\documentclass[aps,prd]{revtex4}
\usepackage{subfigure,graphics} 
\usepackage{flafter} 
\begin{document} 

\title{
Superfluidity in Super-Yang-Mills Theory} 
\author{Ian G. Moss}
\email{ian.moss@ncl.ac.uk}
\affiliation{School of Mathematics and Statistics, University of  
Newcastle Upon Tyne, NE1 7RU, UK}

\date{\today}

%%%%%%%%%%%%%%%%%%%%%%%%%%%%%%%%%%%%%%%%%%%%%%

\begin{abstract}
The AdS/CFT correspondence suggests that there is a point in the phase diagram
of strongly interacting gauge-theory matter where the viscosity
approaches zero. This paper analyses the possibility that this point
represents a superfluid and that the the system near this point in the
phase diagram can be described by a Landau fluid. Superfluid vortices are
constructed and the AdS analogue of vorticity quantisation is described. The
production of vortices in the quark-gluon plasma during heavy ion collisions is
discussed.

\end{abstract}
\pacs{PACS number(s): }

\maketitle
%%%%%%%%%%%%%%%%%%%%%%%%%%%%%%%%%%%%%%%%%%%
\section{introduction}

Experiments with heavy ion collisions at RHIC \cite{Muller:2006ee} have shown
that the quark gluon plasma close to deconfinement can be successfully modelled
using relativistic hydrodynamics.
The properties of the quark gluon plasma should be described by quantum
chromodynamics, but this theory is notoriously difficult to apply in the
strong coupling regime. The AdS/CFT correspondence, developed from superstring
theory, gives a much simpler theoretical framework for certain strongly
coupled gauge theories 
\cite{Maldacena:1997re,Gubser:1998bc,Witten:1998qj}. The prime example is the
${\cal N}=4$ superconformal Yang-Mills theory. In the AdS/CFT correspondence,
the thermal properties of the gauge-theory matter are related to black hole
thermodynamics in 5 dimensions 
\cite{Witten:1998qj,Hawking:1998kw,Chamblin:1999tk,chamblin-1999-60,Caldarelli:1999xj,Gibbons:2004ai}.

Although ${\cal N}=4$ superconformal Yang-Mills theory can only be a toy model
of quantum chromodynamics, it has has been suggested that it might share some
of the features of real quark physics close to the deconfinement phase
transition  
\cite{Nastase:2005rp,Shuryak:2005ia,Janik:2005zt,Gubser:2007zz}.
One influential result which arose in the context of the AdS/CFT correspondence
was the ratio of shear viscosity $\eta$ to entropy density $s$ in
superconformal ${\cal N}=4$ Yang-Mills theory,
\cite{Kovtun:2004de},
\begin{equation}\frac{\eta}{s}={1\over 4\pi}\end{equation}
in natural units ($\hbar=c=k=1$). The results from heavy ion collisions suggest
that the viscosity of the quark gluon plasma might be close to the AdS/CFT
prediction
\cite{Nouicer:2007fk,Romatschke:2007mq}.

We shall be interested here in what happens to gauge theory matter in the limit
$s\to 0$ with non-zero number density, when the viscosity vanishes. The known
mechanism for vanishing viscosity is superfluidity, which is due to the
Bose-Einstein condensation of a composite operator \cite{Feynman:1955}. 
We shall consider the
possibility that this is what happens here, i.e. that as $s\to 0$ the system
becomes a superfluid and can be
described by a condensate.

When the phase of the system lies close to $s=0$, then only part of
the system would be in the condensate and the remainder would be in thermal
excitations. In the
superfluid context, this type of fluid is called a Landau fluid. Under the
AdS/CFT correspondence, we would
expect that two different black holes states should correspond to a single set
of quark gas phase parameters. This is exactly the situation which one often
finds when dealing with black holes in Anti de Sitter space 
(e.g. \cite{Hawking:1983}). The smaller black
hole is usually discarded, but we now look on this hole as representing one of
the two components of the binary fluid.

In order to fix our ideas, consider the thermodynamics of a charged black hole
in Anti-de Sitter space. The black hole corresponds to an equilibrium thermal
field theory on a 3-sphere of radius $l$ \cite{Witten:1998qj}.  The electric
charge of the black hole is related to a conserved quantity in the thermal
system which is similar to baryon number \cite{Chamblin:1999tk} and the
thermodynamic state can be described by a chemical potential $\mu$ and the
temperature $T$. There is a phase transition in the black hole thermodynamics
\cite{Hawking:1983} which corresponds to a quark deconfinement transition
\cite{Witten:1998zw}. 

\begin{center}
\begin{figure}[ht]
\scalebox{0.5}{\includegraphics{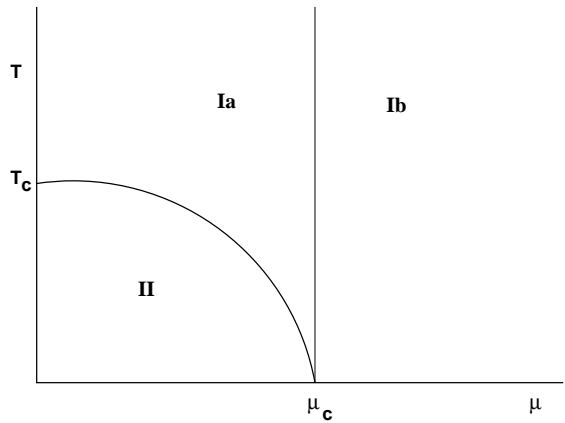}}
\caption{The $(T,\mu)$ phase diagram for Anti de Sitter space. Two charged
black hole solutions exist for parameters in region $Ia$, and one charged
black hole solution for parameters in region $Ib$. Region $I$ is separated
from region II by a second order phase transition. For supersymmetric black
holes, $\mu_c=\sqrt{3}$ in our units where the gravitational coupling strength 
$\kappa_5=1$.}
\label{figph}
\end{figure}
\end{center}

The phase diagram is shown in figure \ref{figph}, where the deconfined phase
has been split into two regions $Ia$ and $Ib$ with
$\mu<\mu_c$ and $\mu\ge\mu_c$ respectively. There are two black hole solutions
for each point in region $Ia$, and one black hole solution for each point in
region $Ib$. If $\mu\to\mu_c$ in region $Ia$, the black hole area of the
smaller hole approaches zero, and according to our hypothesis it represents
the superfluid component of a Landau fluid. The superfluid fraction is
suppressed by its Gibbs free energy. At the end of the phase
transition line lies a point $(T,\mu)=(0,\mu_c)$ where $s\to0$ and the fluid
would become a pure superfluid.

The effective equation for a stationary condensate $\Psi$ takes the form of a
relativistic Gross-Pitaevskii equation (e.g. see
\cite{roberts2000,Kapusta:1989tk}). For
simplicity we shall assume local
interaction terms,
\begin{equation}
-\nabla^2\Psi+g^2|\Psi|^2\Psi+m_B^2\Psi=\mu^2\Psi,
\end{equation}
for some the constants $g$ and $m_B$. The conserved
charge $N_B$ is given by
\begin{equation}
N_B=2\mu\int\Psi^*\Psi\,d\Omega_3\label{nb},
\end{equation}
where $d\Omega_3$ is the volume element on $S_3$.

The AdS/CFT correspondence can lead to some useful information about the
couplings in the Gross-Pitaevskii equation. Consider the ground state with
wave function $\Psi_0=(\mu^2-m_B^2)^{1/2}/g$. When combined with the formula
for the conserved charge (\ref{nb}), this gives an upper bound on the coupling
constant $g^2<4\pi^2\mu^3l^3/N_B$. If the quark theory has gauge group
$SU(N)$, then the radius of the anti de Sitter space is determined by the
formula  $4\pi^2 l^3=N^2$ \cite{Maldacena:1997re}. We can also identify $\mu$
with the value $\mu_c$
obtained through black hole thermodynamics. The upper bound on the coupling
becomes
\begin{equation}
g^2\le N^2\mu_c^3/N_B.
\end{equation}
Both large and small coupling regimes are possible depending on the rank of
the gauge group and the conserved charge. 

Now consider what happens when the fluid rotates. The corresponding charged
 rotating black hole solutions where found by Chong et al \cite{Chong:2005hr}.
The black
holes rotate, but the geometry of the $S_3$ on which the fluid lives remains
unaffected by the rotation \cite{Gibbons:2004ai}. There are two independent
axes of rotation on $S_3$ which lie
on two disconnected circles at $\theta=0$ and $\theta=\pi$ in the Euler angle
parameterisation $(\theta,\psi,\phi)$. The $S_3$ metric in these coordinates
is,
\begin{equation}
ds^2={l^2\over 4}\left(d\theta^2+\sin^2\theta d\phi^2\right)+
{l^2\over 4}\left(d\psi^2+\cos\theta d\phi^2\right).
\end{equation}
A simple class of vortex solutions to the Gross-Pitaevskii equation can be
constructed by taking an anzatz
\begin{equation}\Psi=R(\theta)\,e^{in_a(\phi+\psi)/2+in_B(\phi-\psi)/2}
\label{anz}
\end{equation}
where $n_a$ and $n_B$ are integers. This reduces the Gross-Pitaevskii equation
to an ordinary differential equation in $z=-\cos\theta$,
\begin{equation}
(1-z^2){d^2 R\over dz^2}-2z{dR\over dz}-{n_a^2\over 2(1-z)}R
-{n_b^2\over 2(1+z)}R+\nu(\nu+1)R-
\frac{g^2l^2}{4}R^3=0.
\end{equation}
where $\nu(\nu+1)=(\mu^2-m_b^2)l^2/4$. The boundary conditions are $R=0$ at
$z=-1$ if $n_B\ne 0$ and $R=0$ at $z=1$ if
$n_a\ne 0$.

The vortex solutions represent one or two disconnected vortices at $z=\pm1$.
The
two-vortex solutions only exist when $\nu>1$. The vortices are thin compared to
the radius of the three sphere $l$ for large values of $N$. At more moderate
values of $N$, the vortices have a size comparable to the three sphere and the
modulus of the wave function is always smaller than the ground state value.

\begin{center}
\begin{figure}[ht]
\scalebox{0.7}{\includegraphics{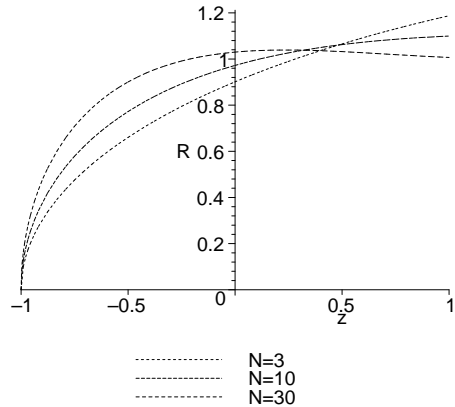}}
\caption{The scaled wave function $gR/\mu$ for a single vortex centred at 
$\theta=0$ or $z=-1$. The thickness of the vortex depends on the rank of the
gauge group $N$ but not on the conserved charge $N_B$. The mass $m_B=0$.}
\label{figpl}
\end{figure}
\end{center}

Solutions of the Gross-Pitaevskii equation can also be described in tems of
relativistic fluid dynamics \cite{roberts2000}. The fluid-flux covector
$(n_B,n_B{\bf u})$ is related to the wave function by
\begin{equation}
n_B=2\mu\Psi^*\Psi,\hbox{        }n_B{\bf u}=
-i(\Psi^*d\Psi-\Psi d\Psi^*).
\end{equation}
The fluid propery of most interest in the case of vortex solutions is the 
circulation around curves $\Gamma$, defined by
\begin{equation}
C=\int_\Gamma {\bf u}.
\end{equation}
For the solution anzatz (\ref{anz}), the circulation is
constant outside the vortex cores, and given by $C_a=2\pi n_a/\mu$ or
$C_b=2\pi n_b/\mu$ for curves around one or the other axis of rotation. In the
non-relativistic
limit, $\mu\approx m_B$ and these reduce to the familiar quantisation of
circulation. 

The quantised quantities are really the angular momenta of the vortex
solutions,
\begin{equation}J_a=n_a\hbox{  and   } J_b=n_b.\end{equation}
These carry accross to the black hole solutions
related to the gauge theory matter through the AdS/CFT correspondence. Quantum
gravity is involved because the angular momenta of the classical black
hole solutions are not quantised. In fact, only the $N\to\infty$ limit
corresponds
to classical gravity, and in this limit the Gross-Pitaevskii
equation breaks down as the thickness of the vortex solutions tends to zero.

There is strong evidence to support the quantisation of black hole angular
momentum. AdS black holes
in 3 dimensions, for example, can be
described by a conformal algebra which implies states of quatised mass and
angular momentum \cite{Birmingham:1998jt}. These play an important role in the
statistical approach to black hole entropy for near-extremal rotating black
holes in 5 dimensions  \cite{Cvetic:1998xh}. The idea here is that the mass,
angular momentum and area can be discussed entirely in terms if the geometry
close to the horizon, which takes the limiting form of a black hole in 3
dimensions. Again, there is a conformal algebra which implies that the angular
momenta along the two axes of rotation are quantised.

The above ideas are closely related to approaches to black hole entropy which
are based on string duality. These relate the properties of the black hole to
an ensemble of $D-$brane states \cite{Strominger:1996sh}. When applied to
extremal rotating black holes in 5 dimensions, one finds that the black hole
angular momenta
are given by quantised $D-$brane charges \cite{Breckenridge:1996is}.

The points in the phase diagram which are associated with zero viscosity are
dense, low temperature states. This seems to be related more to neutron stars
than to heavy ion collisions. The situation when transformed to Minkowki space
is, in fact, rather more interesting. There is a conformal transformation $f$
which
maps $S_3\times R$ to Minkowski
space $R^4$ which takes the time coordinate $t$ and the azimuthal polar
coordinate $\chi$ to the Minkowski light cone coordinates $u$ and $v$,
\begin{eqnarray}
u&=&l\,\tan\left({t\over 2l}-{\chi\over 2}\right)\label{um}\\
v&=&l\,\tan\left({t\over 2l}+{\chi\over 2}\right)\label{vm}.
\end{eqnarray}
This transformation reduces to steriographic projection at $t=0$, but 
in general surfaces of constant time $t$ do not map to surfaces of constant
Minkowski time. The timelike killing vector is
mapped to
\begin{equation}
f_*\partial_t=\frac12\left(P+K\right)
\end{equation}
where $P$ and $K$ are the generators of time translation and timelike conformal
transformations respectively \cite{Horowitz:1998xk}. The conformal symmetry can
be used to relate
operator traces, but thermal states on $S_3$ are related to non-thermal and
non-stationary states in Minkowski space.

Consider operators $\varphi$ and $\varphi'$ with conformal weight 1. The
conformal symmetry relates ensemble averages on $S_3\times R$ to
ensemble averages on Minkowski space,
\begin{equation}
{\rm tr}\left(\rho\,\varphi(x)\varphi(x')\right)
=\Omega(x)\Omega(x'){\rm tr}\left(\rho'\,f^*\varphi'(x)f^*\varphi'(x')\right),
\end{equation}
where $\Omega$ is the conformal factor. The thermal states with
$\rho=e^{-\beta H}$ are related to ensemble
averages in Minkowski space with a density matrix
\begin{equation}
\rho'=e^{-\beta(H'+K')/2}
\end{equation}
where $H'$ and $K'$ are the Minkowski space Hamiltonian and generator of
conformal transformations respectively. The new density matrix does not commute
with the Hamiltonian, or with any other Poincar\'e group generator.

\begin{center}
\begin{figure}[ht]
\scalebox{0.5}{\includegraphics{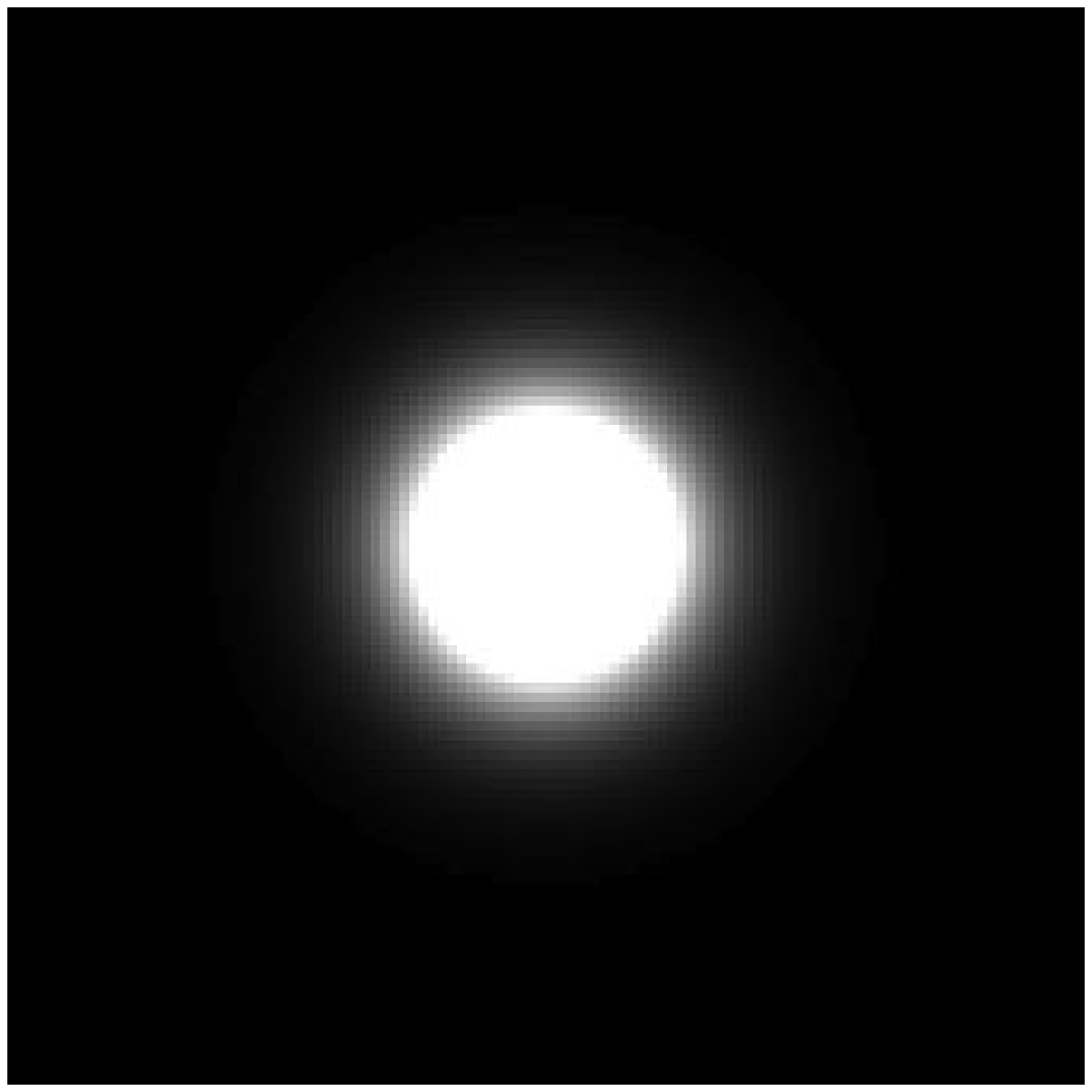}}
\scalebox{0.5}{\includegraphics{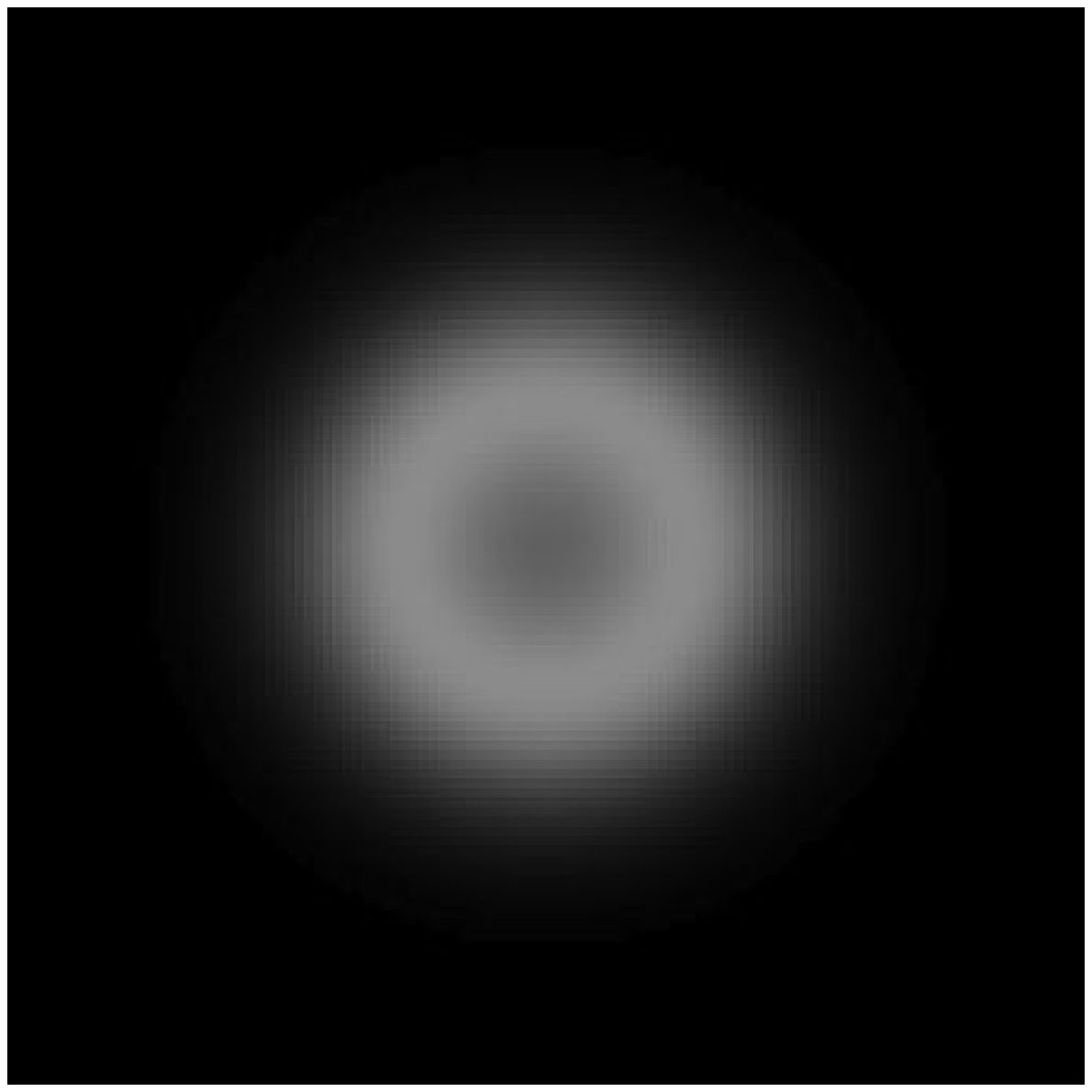}}
\caption{The white sphere shows the spatial distribution of the energy density
of the gauge matter fluid at the initial time $t=0$ when it is at rest and at
a later time $t=l$ as it expands. }
\label{figsp}
\end{figure}
\end{center}

The conformal symmetry allows the energy density of the gauge-theory matter in
Minkowski space to be expressed in terms of the conformal factor $\Omega$ and
its derivatives. The result is shown in figure \ref{figsp}. The Minkowski
space ensemble represents a ball of
fluid which starts from rest and then expands and dissipates away.

Even though the Minkowski system is non-thermal, it is possible to define a
local temperature by examining the KMS condition of periodicity in imaginary
time when the points $x$ and $x'$ lie close to the centre of the ball of fluid.
The effective inverse temperature $\beta'$ at the centre is given in an
elementary way by substituting $t=i\beta$ in eq. (\ref{um}),
\begin{equation}
\beta'=l\,\tanh\left({\beta\over 2l}\right)
\end{equation}
The $T\to0$ limit which we associated with the superfluid state corresponds to
a temperature $T'\to l^{-1}$ at the centre of the ball of fluid. The
image of the Landau fluid in Minkowski space is therefore effectively at or
above the temperature which we associated with a deconfinement phase
transition.

\begin{center}
\begin{figure}[ht]
\scalebox{0.5}{\includegraphics{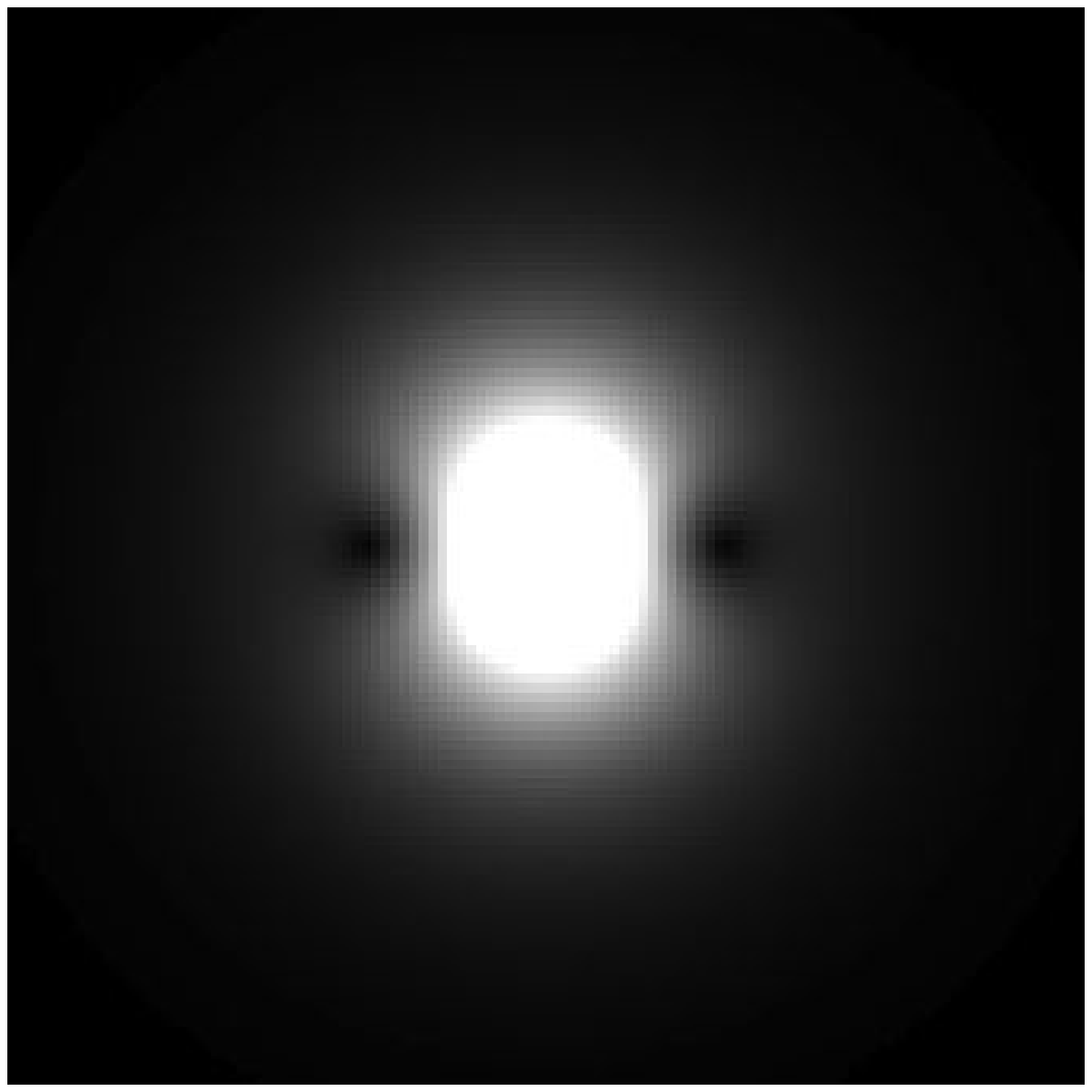}}
\scalebox{0.5}{\includegraphics{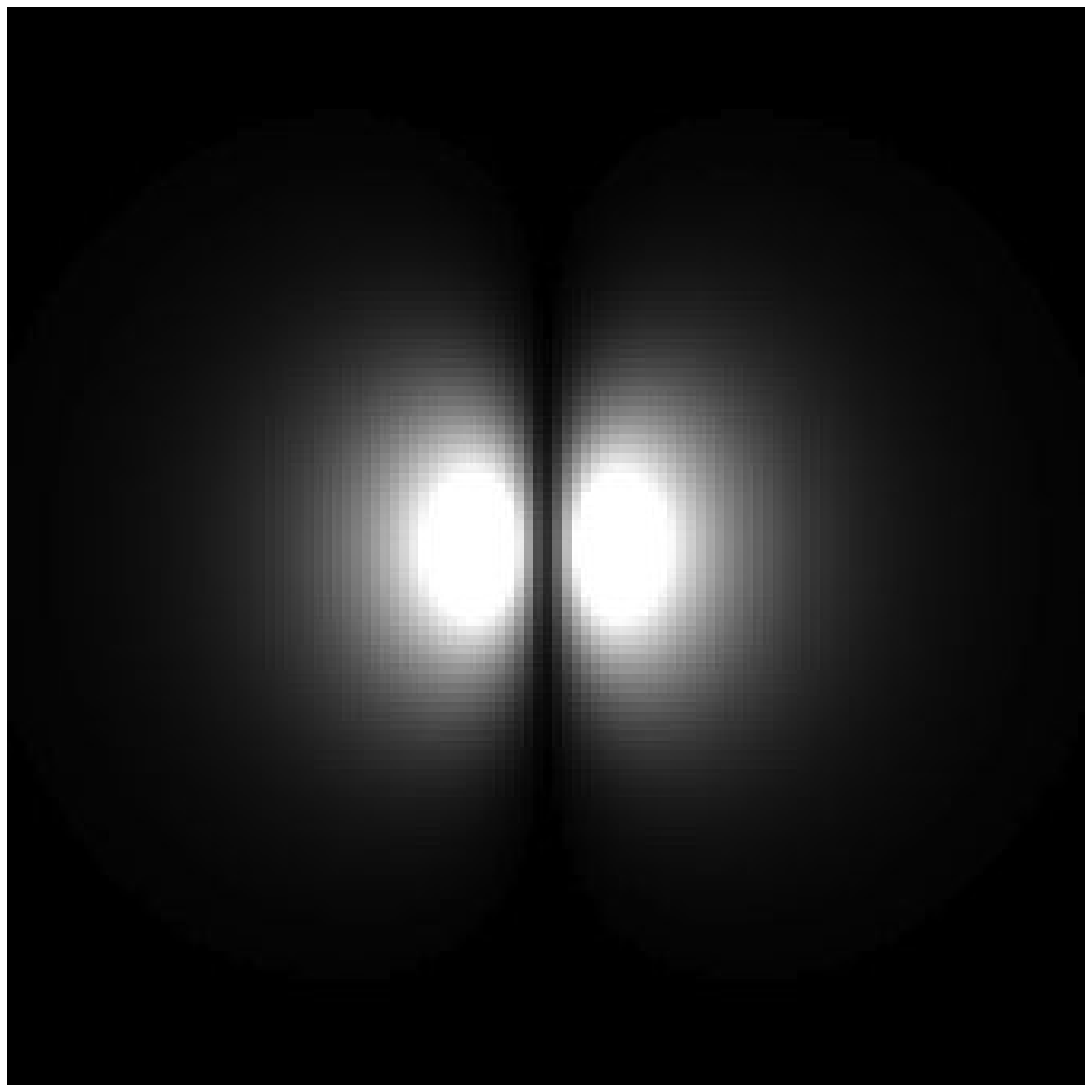}}
\caption{The spatial distribution of the number density of the gauge matter
fluid is shown on a vertical slice through a horizontal ring vortex (left) and
a vertical line vortex (right). The fluid is ejected from the vortex core.
These figures
correspond to the middle curve in figure \ref{figpl}.}
\label{figri}
\end{figure}
\end{center}

The vortex solutions can also be mapped to flat space by the conformal
transformation (\ref{um}) and (\ref{vm}). The images of the solutions are
circular ring vortices or line vortices. Double vortex solutions are mapped to
interlinked vortices which lie in two orthogonal planes. Figure \ref{figri}
shows a section through a ring vortex and a line vortex at the initial
Minkowski time.

We turn finally to the question of whether these superfluid concepts can tell
us anything about the quark-gluon plasma. In particular, can we produce
superfluid vortices in heavy ion collisions? In relativistic heavy ion
collisions, Lorentz contraction of the nuclei results is a planar collision
geometry. After a short time has elapsed, thermal behaviour sets in and the
energy becomes spread over an increasingly spherical region 
\cite{Shuryak:2003xe}. The
baryon chemical potential in ultrarelativistic collisions 
\cite{Arsene:2004fa,Adcox:2004mh,Back:2004je,Adams:2005dq} 
$\mu\approx 10\hbox{MeV}$, appears to be far smaller compared to the
temperature  of the
quark-gluon plasma $T\approx 170\hbox{MeV}$ than we would need for a Landau
fluid description.

The picture of gauge matter obtained from spherical black holes differs from
those models in which gauge-theory matter is modelled by an evolving black
brane \cite{Janik:2005zt,Shuryak:2005ia,Nastase:2005rp}. These are based on
planar geometry, but it would be interesting to extend these ideas to
base the evolution on spherical background geometry using conformal mappings of
the type discussed here.

\acknowledgements
I would like to thank Paul Mackay for assistance with the vortex equations.

%%%%%%%%%%%%%%%%%%%%%%%%%%%%%%%%%%%%%%%%%%%%%
\bibliography{paper.bib}

\end{document}